What Should be the Role of Field Energy in Introductory Physics Courses?

Robert C. Hilborn, *American Association of Physics Teachers, College Park, MD 20740*

*A Framework for K-12 Science Education: Practices, Crosscutting Concepts, and Core Ideas* gives the concept of field energy a prominent role in the physical sciences sections of its recommendations for K-12 science education. I examine what *A Framework* suggests for the role of field energy and point out that, given the ambiguities and complexities associated with field energy, a traditional approach focusing on potential (configuration) energy is more appropriate for introductory physics in secondary schools, colleges, and universities.

## I. INTRODUCTION

In the recently released National Research Council report *A Framework for K-12 Science Education: Practices, Crosscutting Concepts, and Core Ideas*[1], the concept of field energy (particularly electromagnetic field energy) plays a prominent role, an unusual departure from traditional treatments of energy in K-12 science education. The authors of *A Framework* focus on field energy in an (apparent) attempt to provide a common concept for describing the structure of atoms, molecules, bulk matter and electromagnetic radiation. They also suggest using gravitational field energy to talk about the gravitational interaction of objects. Given the well-known mathematical ambiguities and conceptual complexities associated with field energy, I argue here that K-12 students are better served by focusing on potential (configuration) energy rather than field energy. Although *A Framework* focuses on K-12 science education, my conclusions also apply to the use of field energy in introductory science courses in higher education.

The arguments in this paper are not new: the difficulties associated with electromagnetic field energy and models of charges have been known for over 100 years[2]. They were also forcefully stated by Feynman in his famous *Lectures*[3] in the 1960s and are discussed in detail in most upper-level textbooks on electrodynamics[4].

I claim that the use of field energy in K-12 science education is undesirable for several reasons: (1) there is no easily observable evidence for field energy (except as noted below). Consequently, students would have to accept the notion only on the basis of authority—counter to the science practice, articulated in *A Framework*, that students ought to engage in explanation from evidence[1]. (2) Even if the students were to accept the concept of field energy, there is almost nothing they can do with that concept quantitatively. Even the simplest case of electric field energy—that due to two charged objects—leads the students (and their teachers) into a swamp of infinities and complex calculations. (3) The *Framework* document makes several claims about the behavior of field energy that are not consistent with what we know about the structure of matter modeled as a collection of charged objects (electrons and nuclei for ordinary matter) whose behavior is described by quantum mechanics. Those statements, if not completely erroneous, apply only in highly restricted situations. Given these complexities, I argue that field energy, though an important concept in contemporary physics, is a relatively sophisticated one that ought to come later rather than sooner in students' learning of science.

## II. WHY FIELD ENERGY?

*A Framework* begins its discussion of field energy on page 121:

> Energy is best understood at the microscopic scale, at which it can be modeled as either motions of particles or as stored in force fields (electric, magnetic, gravitational) that mediate interactions between particles….Electric and magnetic fields also contain energy; any change in the relative positions of charged objects (or in the positions or orientations of magnets) changes the fields between them and thus the amount of energy stored in those fields.

This passage indicates that the authors of *A Framework* are using field energy as part of a statement about conservation of energy. There are, however, several problems with this passage. First, how might one justify the assertion that "energy is best understood at the microscopic scale"? Is that statement true for the kinetic energy of a truck speeding down a highway? Is the energy stored in the gravitational field associated with the Sun and the planets microscopic or macroscopic? Next, the term "force field" can be misleading. Certainly electric, magnetic, and gravitational forces are usefully explained by and calculated from fields, but the field values, as usually expressed in physics, are not values of forces. (See the Appendix to this paper for a further discussion of "force field.") Also, the "either…or" in the passage will inevitably be taken to be an "exclusive or": energy is either kinetic or field energy—not both. In reality, both forms of energy are always present. Finally, the phrase "the fields between them" is rather vague and is in fact contrary to the fact, discussed in detail below, that only field energy integrated over all space is physically significant. I suggest that a better formulation of the first sentence is "At the microscopic scale, energy can be modeled as a combination of energy associated with the motion of particles (kinetic energy) and the energy stored in fields (electric, magnetic, and gravitational) that mediate interactions among particles."

This focus on field energy also appears in the College Board College Readiness Standards[5], where it is promoted as a concept for all K-12 students. Ironically, field energy is not mentioned at all in the Advanced Placement Physics 1 and Physics 2 description[6] nor in the AP Physics C (calculus-based) course description[7].

Most introductory physics and physical science textbooks and courses use the concept of configuration (potential[8]) energy when talking about conservation of energy. It appears that the authors of *A Framework* and the *College Board Standards* seek to replace the notion of configuration energy with the concept of field energy. As an example of why field energy might be useful, consider the head-on collision of two low-friction carts with magnetic "bumpers" (with like magnetic poles facing each other giving a repulsive force between the carts). As the carts approach each other, the kinetic energy of the carts decreases as the carts slow down and then increases again as the carts travel apart. While the carts are slowing down, the kinetic energy must be stored somewhere, the usual reasoning goes. *A Framework* encourages us to say that the energy is temporarily stored in the magnetic fields associated with the two magnets. The more traditional approach would state that the energy is stored as configuration (potential) energy associated with the relative position of the two carts. In this situation, the field energy change is, to a good approximation[9], the same as the configuration energy change. So, either description would be fine. But we need to consider more than just one simple example to weigh the relative merits of field energy compared to those of configuration energy.

Let us now turn our attention to the physics of field energy to see whether the equivalence between field energy changes and potential energy changes is true in general and to explore the complexities mentioned previously.

## III. EVIDENCE FOR FIELD ENERGY

The magnetic bumper example indicates that in some situations we could use either configuration energy or field energy to understand the changes in kinetic energy. As with most static or quasi-static situations where dissipative forces can be neglected, the configuration energy approach is used in traditional physics instruction for gravitational, electrical, and elastic interactions. Field energy is certainly not the only way to describe such situations and in current K-12 science instruction, it rarely appears.

Are there situations where field energy is necessary? Field energy becomes essential in situations where the fields are dynamic and changing rapidly in time, leading to the possibilities of delayed interactions and radiation[10]. In the case of radiation, parts of the fields are in a sense "cut loose" from their sources and can propagate on their own, carrying energy (and momentum and angular momentum).

Let us use that idea to formulate a line of reasoning that justifies thinking about energy stored in electromagnetic fields: Students can observe that light carries energy (heating up objects, causing fluorescence, activating a cell phone, etc.). Knowing that electric fields exert forces on charges and given that materials are composed of charged objects, students can legitimately conclude that light has an electric field associated with it. Since light can carry energy through otherwise empty space (e.g. light from the Sun), it is reasonable to conclude that the energy can be associated with the electric field (and, by generalization, the magnetic field). I note that this line of reasoning requires a fairly sophisticated background: knowing that electric fields exert forces on charged objects, that accelerating charges produce electromagnetic waves, and that matter is made of electrically charged objects that interact primarily through electromagnetic fields. Nevertheless, the conclusions about field energy are based on observations. It is worth pointing out that we don't need the concept of field energy to understand *that* light (and other electromagnetic waves) transfer energy. But field energy, admittedly, gives us an explanation of *how* electromagnetic waves transfer energy.

## IV. OFF TO INFINITY

The second point about ambiguities and complexities associated with field energy needs some quantitative and formal statements. For the sake of simplicity, I focus on just electric field energy. As is well known, the electric field energy per unit volume at a particular location is proportional to the square of the electric field $E^2$ at that location. If several charges are present, the relevant electric field is the net field obtained by adding up the fields (as vectors) from all of the charges. To get the total field energy, which is the important concept for the ideas *A Framework* is promoting, we then need to integrate the energy per unit volume over all space.

Though the process sounds relatively straightforward, there are several difficulties in implementing this approach. The first difficulty is apparent even if we have only one point charge present. The magnitude of the electric field for a point charge $q$ located a distance $r$ from the "observation point" in question is given by

$$E(r) = k\frac{q}{r^2}, \tag{1}$$

where the constant *k* takes care of the units used for field, charge, and distance. The well-known problem is that the total field energy associated with the point charge is infinite. The offending location is at the point charge where the field itself becomes infinite. Any spatial volume containing the point charge will have infinite electric field energy associated with it.

We can rescue the situation by noting that we are primarily concerned with <u>changes</u> in field energy when charges change location with respect to each other. Let's consider the simple case of two point charges *a* and *b*. The square of the net field for the two point charges is given by the expression $\left(\vec{E}_a + \vec{E}_b\right)^2 = E_a^2 + E_b^2 + 2\vec{E}_a \cdot \vec{E}_b$. I shall call the last term "interaction field energy density" because it is significant only when the two charges are close enough for their interactions to be important.

What happens to the field energy is this case? As was noted before, the first two terms each lead to infinite field energy when integrated over all space. But those results are unaffected by changes in the relative position of the two charges and one might argue (leaving general relativity aside for the moment) that only changes in energy are important. It turns out that this procedure works, even though mathematicians cringe when we subtract infinity from infinity. To get the interaction field energy, we subtract the infinite field energy associated with the two charges when they are very far apart from the infinite field energy (because of the $E_a^2$ and $E_b^2$ terms) when the $\vec{E}_a \cdot \vec{E}_b = E_a E_b \cos\theta_{ab}$ is significant, and we end up with a finite result. Note that the $\vec{E}_a \cdot \vec{E}_b$ term can be positive in some regions of space (where $0 \leq \theta_{ab} < \pi/2$) and negative in others (where $\pi/2 < \theta_{ab} \leq \pi$). It is not obvious, without detailed calculations, that the <u>total</u> interaction field energy decreases when opposite-sign charged particles attract and when like-sign charges repel as asserted (correctly) in *A Framework* :

> Force fields (gravitational, electric, and magnetic) contain energy and can transmit energy across space from one object to another. When two objects interacting through a force field change relative position, the energy stored in the force field is changed. Each force between the two interacting objects acts in the direction such that motion in that direction would reduce the energy in the force field between the objects. (p. 127)

The Appendix of this paper shows how this association works for the simple case of two point charges. An easy generalization to a system of many point charges shows that for electrostatic situations, the interaction field energy changes are equivalent to the electric configuration (potential) energy changes, although the latter are much easier to calculate. Finding the change in field energy directly for more complex systems quickly becomes intractable.

## V. A LONG HISTORY

As mentioned previously, the difficulties associated with field energy and point charge models have been long known. The essential point is that the concept of field energy is inconsistent with point charge models. This inconsistency is discussed in *The Feynman Lectures*[3]:

> The difficulty we speak of is associated with the concepts of electromagnetic momentum and energy, when applied to the electron or any charged particle. The concepts of simple [point-like] charged particles and the electromagnetic field are in some way inconsistent. (Vol. II, p. 28-1)

The same issues show up in quantum electrodynamics[11], the quantum theory of the electromagnetic field and its interactions with charged particles. Weinberg, in his masterful *Quantum Theory of Fields*[12], writes

> Earlier experience with classical electron theory provided a warning that a point electron will have infinite electromagnetic self-mass; this mass is $e^2/6\pi ac^2$ for a surface distribution of charge with radius *a*, and therefore blows up for $a \to 0$. Disappointingly this problem appeared with even greater severity in the early days of quantum field theory, and although greatly ameliorated by subsequent improvements in the theory, it remains with us to the present day. (Vol I, p.31)

The infinities in quantum field theory and how they are avoided are discussed in some detail Ref. 13.

## VI. FIELD ENERGY AND THE STABILITY OF MATTER

We now turn to the third issue, which is more subtle and includes some scientifically erroneous statements in the *Framework* document. *A Framework*[1] includes claims about how the stable forms of matter are related to electromagnetic field energy. On pages 109 and 239, we find

> Stable forms of matter are those in which the electric and magnetic field energy is minimized.

A somewhat expanded statement is found on page 121:

> Matter in a stable form minimizes the stored energy in the electric and magnetic fields within it; this defines the equilibrium positions and spacing of the atomic nuclei in a molecule or an extended solid and the form of their combined electron charge distributions (e.g., chemical bonds, metals).

Although these statements sound plausible, there are several problems lurking here. In the second statement, it is not clear what the phrase "the stored energy in the electric and magnetic fields within it" means. What defines "within it"? The only mathematically and conceptually meaningful electromagnetic field energy, as mentioned previously and demonstrated explicitly in the Appendix of this paper, is that obtained by integrating the interacting field term over all space. For field energy, there is no way to define a "within" and "outside." The fields are present everywhere and all locations must be included to get the total energy stored in the field. Perhaps the authors mean that one needs to consider the fields due to the constituents of the system, not fields due to sources outside the system. If so, they should make that point explicit.

The first statement about the stable configuration of matter is approximately true[14] when thinking about how atoms (or ions), as already constructed objects, arrange themselves in molecules and in bulk matter. However, it is not true in general. To illustrate the difficulty, let's look at a very simple case: the hydrogen atom—the bound state of one electron and one proton. In classical (pre-quantum) mechanics, an electron orbiting the proton will radiate electromagnetic energy and will spiral in towards the proton. The stable state occurs when the electron and proton come together at which point the electromagnetic field energy is minimized because the fields become very small as the charges overlap, giving zero net charge with no or almost no separation between the positive and negative charges (the details obviously depend on the model of the structures of the proton and electron).

However, we know that for the actual stable, lowest energy state of a hydrogen atom, the electron is on average about 0.05 nm from the proton—obviously not the configuration for the minimum of electromagnetic field energy. Nowadays, we understand how this comes about from quantum theory: the Heisenberg Uncertainly Principle requires that a decrease in the average distance between the electron and proton requires a larger (internal) kinetic energy. The lowest energy (ground) state of the hydrogen atom can be viewed as a compromise between lowering the configuration energy (and the corresponding interaction field energy) as the distance between the electron and the proton decreases and the corresponding increase in kinetic energy required by the Uncertainty Principle.

For more complex atoms (with more electrons) and for bulk matter, the situation is further complicated by the requirements of the Pauli Exclusion Principle: no two electrons can occupy the same quantum state. So the stable states of the system must satisfy both the Uncertainty Principle and the Exclusion Principle. We conclude that minimizing the electromagnetic field energy of a system does not determine the stable form (configuration) of matter. The laws of quantum mechanics play a crucial role. In fact, quantum calculations for molecules and solids typically use either the Coulomb potential between charges or some "effective" potential (averaging over electron positions) to find the system energies as a function of inter-nuclear distances. The ground state energy of the system is found by minimizing the energy associated with those potentials, subject to the rules of quantum mechanics[15]. In effect, the (infinite) field energy associated with assembling the individual charges and the field energies associated with building the atoms from those charges are ignored.

How did such an erroneous statement arrive in *A Framework*? It is true that the effective configuration energy given as a function of inter-nuclear distance for diatomic molecules, for example, does have a minimum at the (stable) equilibrium (ground) state of the molecule. It seems that the authors of these passages about molecules and solids in *A Framework* have assumed a simple relationship between a minimum in the configuration energy (usually determined empirically or theoretically from quantum mechanics) and a minimum in the electromagnetic field energy. As I have argued, if there is a relationship, it is one between the interaction field energy and the electrostatic potential energy, a rather limited relationship.

## VII. FOCUS ON POTENTIAL (CONFIGURATION) ENERGY

These difficulties associated with field energy strongly suggest that a more productive pedagogical strategy for K-12 students is to focus on energy stored in the configuration of objects, that is, on what is commonly called potential energy. In fact, that shift in focus is hinted at on page 121 in *A Framework*:

> Energy stored in fields within a system can also be described as potential energy. For any system where the stored energy depends only on the spatial configuration of the system and not on its history, potential energy is a useful concept (e.g., a massive object above Earth's surface, a compressed or stretched spring). It is defined as a difference in energy compared to some arbitrary reference configuration of a system. For example, lifting an object increases the stored energy in the gravitational field between that object and Earth (gravitational potential energy).

Unfortunately, the first sentence in this quotation is wrong as a general statement. In fact, it is valid only if radiation and velocity-dependent effects can be neglected. A correct restatement of the first sentence would go something like this: "Energy stored in fields in a static or quasi-static system (small relative velocities and small accelerations) can also be described as potential energy." Potential energy is no longer useful if the components of the system have sufficiently large accelerations to produce significant radiation. Then, as mentioned previously, we must use the concept of field energy. Note that the other sentences quoted above indicate that configuration energy has uses in situations (such as springs, drumheads, and elastic beams) where there is no easily identifiable field energy.

Let's examine the last sentence in detail since gravitational potential energy is often students' first encounter with the notion of potential energy. Indeed it is true that lifting an object above the surface of Earth increases the (interaction) field energy associated with Earth and the object. More commonly, students learn that the change in vertical position of the object relative to Earth increases the configuration (potential) energy associated with the Earth-object system. In this case, the two energy changes are the same, but it is extremely difficulty to calculate the change in gravitational field energy when an object with mass $m$ is raised above the surface of the Earth. On the other hand, all students quickly learn that the change in gravitational potential energy of the system when the vertical position of the object is changed by $h$ is simply $mgh$ since that change can be readily calculated from the work done by the gravitational force that Earth exerts on the object as the object's vertical position changes. Although it takes time for students to develop a deep conceptual understanding of configuration energy, they can readily use the expression for the change in gravitational potential energy to predict the motion of roller coasters and objects in free fall, so that over time, they have evidence that there may be something to the notions of configuration energy and conservation of energy.

## VIII. CONCLUSION

The arguments presented here suggest that for most situations, it would be better to focus on configuration energy rather than field energy. Configuration energy hides the infinities associated with the field energy because it focuses on the energy needed to set up a configuration of objects relative to some reference configuration, not on the energy associated with building the objects themselves. Also, the configuration energy is generally easier to calculate since it involves the scalar potential energy functions rather than vector quantities that describe the fields. Furthermore, the configuration energy formulation has the advantage that it is expressed directly in terms of the position coordinates of the components of the system. The time derivatives of those components give the velocities and accelerations of the components, making the comparison of configuration energy and kinetic energy more transparent. In addition, for a system of point-like objects, the configuration energy function does not require any integration over spatial domains just a sum over the locations of the objects. Finally, I note that the concept of configuration energy can also be used for elastic material systems such as springs and strings where the concept of field is not useful and can thus unify treatments of microscopic and macroscopic systems.

There are many problematic and, in some cases, erroneous statements about field energy in *A Framework*. Given the complexities of field energies, there is little gained in thinking about field energy unless one is dealing with radiation. In almost all other cases, configuration energy is a more useful concept both scientifically and pedagogically.

If *A Framework* is widely adopted, most students and teachers will repeat the words about energy being stored in fields. There is nothing fundamentally wrong there, but they will find themselves in the awkward position of using those words without being able to cite evidence for their correctness, nor will they be able to use those words to understand phenomena in the world or to predict the behavior of systems in the world. More importantly, I argue that documents which purport to set a framework and performance expectations for students across the country ought to get the science right. With reasonable thought and care, language can be constructed that both gets the science right and is pedagogically and developmentally appropriate for K-12 students (and beyond). There is also a political danger in having ambiguous and sometimes erroneous statements in these documents: An expert scientist or engineer will seize upon these deficiencies and then argue that the entire science standards enterprise must be suspect if the authors can't even get the basic science correct, thereby providing ammunition for those who are opposed to national education standards.

**Appendix**

In this appendix I address some of the more formal aspects of field energy to confirm the statements made in the main part of the paper.

**1. Force Field**

First, I address the issue of "force field," which appears in *A Framework* in several passages, one of which is cited in the main part of this paper. The notion of force field is a legitimate concept but one that is rarely used, except in an informal, colloquial sense, in introductory science courses.

A force field[16] presumably describes the forces exerted on some object (by other objects) as a function of spatial position. Let's take the simplest case of two objects *a* and *b*. The force that *a* exerts on *b* can be written as $\vec{F}_{ab}(\vec{r}_{ab})$, where $\vec{r}_{ab}$ is the position vector for *b* relative to *a*. (I assume that the force that *a* exerts on *b* depends only on relative position, not on velocity, or acceleration.) Of course, there is also a force field for the force that *b* exerts on *a* as a function of position. The notion of force field says nothing about how object *a* interacts with object *b* except that the interaction can occur at a distance; the objects do not need to be in contact. The interaction could be explained by some "action at a distance" mechanism. A field as a mediator of long-range interactions is a separate issue.

It is conventional (and very useful) to talk about the interaction between *a* and *b* in terms of a field that is associated with *a* and how that field at the location of *b* interacts with object *b* to give the force that *a* exerts on *b*. The field associated with *a* extends through all space and provides a spatially local explanation of the interaction between *a* and *b*. Similarly, we say that *b* has a field associated with it that produces the force exerted on *a*. For the case of electrically charged objects, we say that charge *a* has an electric field $\vec{E}_a(\vec{r})$ associated with it and the force exerted on charge *b* at location $\vec{r}_b$ is given by $q_b \vec{E}_a(\vec{r}_b)$, in other words the electric field gives the force per unit charge at that location in space.

As *A Framework* emphasizes, electric, magnetic, and gravitational fields store energy. However, the "force field" as described above is not directly related to the energy stored in the electric, magnetic, or gravitational fields. The energy (per unit volume) stored in an electric field at a specified location is proportional to the square of the net electric field at that location $\left(\vec{E}_a(\vec{r}) + \vec{E}_b(\vec{r}) + ...\right)^2$; that is, we need to add up at that location the electric fields from all of the relevant charges. If we wanted to use the term "force field," we could say that $\left(\vec{E}_a(\vec{r}) + \vec{E}_b(\vec{r}) + ...\right)$ is the force field for a charge with $q = 1$, assuming that the presence of that $q = 1$ charge does not affect the positions of the other charges in the system. But, from this point of view, the term force field does not add anything to the discussion.

The conceptual difficulty is that there are several "force fields" present: the force that charge *a* exerts on charge *b*, the net force exerted on charge *b* (for example), and the $q = 1$ (test charge) force field. It is the latter that is related to the conventional definition of field energy. To avoid ambiguity, it would be better to be specific and say electric field, magnetic field or gravitational field, or whichever combination is appropriate, rather than the general (and ambiguous) term force field[17].

## 2. Field Energy Formalism

I now show that the electric field energy associated with a point charge is infinite. Using the expression for the electric field of a point charge $E = kq/r^2$ in the expression for the field energy and employing spherical coordinates yields

$$W_a = \tfrac{1}{2}\varepsilon_0 \int_0^{2\pi} d\phi \int_0^{\pi} \sin\theta\, d\theta \int_0^{\infty} dr\, \frac{r^2 k^2 q_a^2}{r^4}, \qquad (2)$$

where I have used SI units, $W_a$ is the electric field energy associated with charge $q_a$. The result from integrating over $r$ (before taking the limits) is proportional to $1/r$ and goes to infinity at the lower limit.

Next I show how to evaluate the interaction energy stored in the fields of two static point charges. (A similar calculation for the case of two equal charges was done in Ref. 18, whose authors call this term "the energy of interaction.") Since I am concerned with how the field energy depends on the distance between the charges, I ignore the (infinite) field energy associated with each of the two charges individually since that energy is independent of separation between the two objects for point charge models.

The total interaction field energy for the two point charges $q_a$ and $q_b$ is given by the integral

$$W_{ab} = \frac{1}{2}\varepsilon_0 \int_{\text{all space}} 2\vec{E}_a \cdot \vec{E}_b\, dV. \qquad (3)$$

To evaluate this integral, it is advantageous to use a spherical coordinate system with the origin at the location of point charge $a$ as shown in Fig. 1.

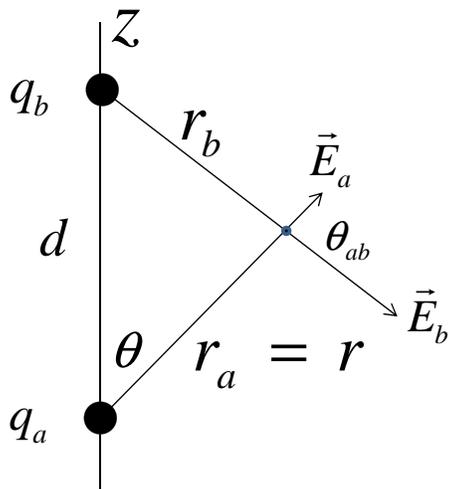

**Figure 1.** The geometry for the two-point-charge model. The $z$ axis is chosen to be along the line joining the two charges $q_a$ and $q_b$. The origin is chosen to be at the location of charge $q_a$. $d$ is the distance between the two charges.

Since the system is cylindrically symmetric about the line joining the two charges, it is helpful to align the $z$ axis of the coordinate system along that line. The integral for the total interaction field energy is

$$W_{ab} = \varepsilon_0 \int_0^{2\pi} d\phi \int_0^{\pi} \sin\theta\, d\theta \int_0^{\infty} \frac{k^2 q_a q_b \cos\theta_{ab}\, r^2 dr}{r^2 r_b^2}, \qquad (4)$$

where $\theta_{ab}$ is the smaller of two angles between $\vec{E}_a$ and $\vec{E}_b$, and I have made use of the volume element in spherical coordinates $dV = r^2 \sin\theta \, dr \, d\theta \, d\phi$. With the aid of the Law of Cosines, $r_b$ and $\cos\theta_{ab}$ can be expressed in terms of $r$ and $\theta$:

$$\cos\theta_{ab} = \frac{r^2 + r_b^2 - d^2}{2 r r_b}, \tag{5}$$

where

$$r_b^2 = r^2 + d^2 - 2rd\cos\theta. \tag{6}$$

Using these results, I find that the $r$ integral becomes

$$\int_0^\infty \frac{r - d\cos\theta}{\left[r^2 + R^2 - 2rd\cos\theta\right]^{3/2}} dr. \tag{7}$$

Note that the integrand in Eq. (7) can be positive or negative, indicating that with both like–sign charges and unlike-sign charges, the local interaction field energy density is in some places positive and other places negative. The *total* interaction field energy is negative for two unlike charges and positive for two like charges, as is expected from electrostatic potential energy considerations.

The integral in Eq. (7) can be evaluated through the use of trigonometric substitutions, integral tables, or symbolic manipulation packages such as *Mathematica*. The result is just $1/d$. The $\theta$ and $\phi$ integrals are easy and yield $4\pi$. Assembling all the pieces, I find that the total interaction field energy is given by

$$W_{ab} = \frac{1}{4\pi\varepsilon_0} \frac{q_a q_b}{d}, \tag{8}$$

just the expression that would have been found if I had used $V_a = \frac{1}{4\pi\varepsilon_0}\frac{q_a}{r}$ for the electrostatic potential due to point charge $q_a$ (assuming as usual that the electrostatic potential goes to zero very far from the point charge).

I now point out an interesting fact about where the interaction field energy density is positive and where it is negative. The dot product $\vec{E}_a \cdot \vec{E}_b = E_a E_b \cos\theta_{ab}$ tells us that the interaction energy density is positive when $0 \leq \theta_{ab} < \pi/2$ and negative when $\pi/2 < \theta_{ab} \leq \pi$, as mentioned in the main text of this paper. The condition $\theta_{ab} = \pi/2$ gives the boundary between the two regions. An examination of the integrand in Eq. (7) indicates this condition is equivalent to $r = d\cos\theta$, with $0 \leq \theta \leq \pi/2$. Figure 2 shows the curve (dashed line) determined by this equation, which is just a circle with radius $d/2$. Because of the system's cylindrical symmetry about the line joining the two charges, the boundary between positive and negative interaction field energy densities is actually a figure of revolution (in this case a sphere) about that line. For two like-sign charges, the interaction energy density inside the sphere is negative, while it is positive outside the sphere. The signs are reversed for opposite-sign charges. As mentioned previously, for like-sign charges, the total interaction field energy is positive; for opposite charges, the total interaction field energy is negative. In both cases, the contribution of the region outside the sphere dominates the total interaction field energy.

Note that the shape of the boundary is independent of the magnitude and sign of the two charges. The boundary is determined simply by the geometry of the fields, which are directed radially towards or away from the individual charges. To show how this works, I note that the boundary can also be specified as that region in space where $\vec{E}_a \cdot \vec{E}_b$ is zero, that is, where the two field vectors are perpendicular. The dashed line circle satisfies that condition because the angle between the two field vectors is the "inscribed angle" for the circle (in Figure 2) and by the famous Central Angle Theorem, the inscribed angle $\theta_{ab}$ is twice the "central angle," which in this case is equal to $\pi$. So, I find that $\theta_{ab} = \pi/2$ everywhere on the circle. Note that the location of the $q_a$ is a singular point where the direction of the field due to $q_a$ is not defined. (An analogous statement holds for the location of $q_b$.) The convention is to use the tangent to the circle as the direction of the field due to that charge at that location.

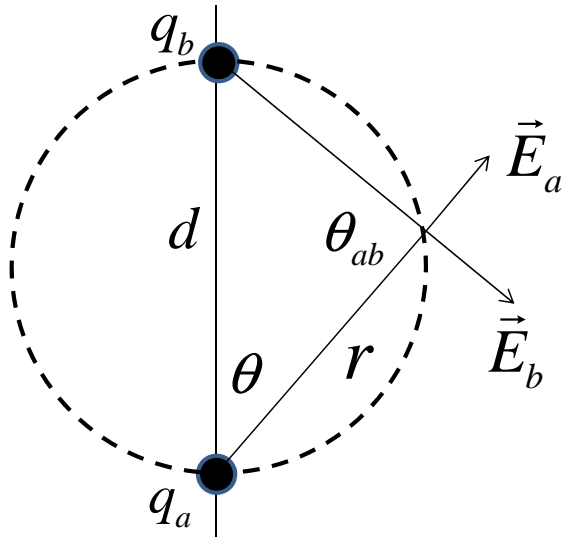

**Figure 2**. The dashed line indicates the boundary between regions where the interaction field energy density is positive and where it is negative. For two like-sign charges, the interaction field energy density is negative inside the sphere and positive outside.

The same kind of analysis can be used to give more details about where the interaction field energy is located. Intuition might lead us to conclude that the interaction field energy is concentrated between the two charges, where, in some loose sense, the fields are strongest. In fact, the ratio of the field energy inside the dashed line sphere in Fig. 2 to the field energy outside the sphere is given by $a/(1-a) = -0.2203...$, where $a = \frac{1}{2}(1 - \pi/2)$. That result tells us that most of the interaction field energy is found outside the sphere, not "between the objects" as *A Framework* asserts in the quotations cited in the main text of this paper or as one's (naïve) intuition might lead one to believe.

This formulation can also be used to show quantitatively how the field energy varies with distance. With the origin at charge $q_a$ (if the two charges are not the same, there is no obvious symmetry advantage to using any other origin), the field energy for $r > d$ goes as

$$W_{ab} = \frac{q_a q_b}{4\pi\varepsilon_0}\left(\frac{1}{d} - \frac{1}{r}\right) \tag{9}$$

For a sphere of radius $r = d$ centered at $q_a$, the net interaction energy is zero. The region outside that sphere has, in a sense, all of the net interaction field energy. To enclose 90% of the interaction field energy, we need a sphere of radius $r = 10\,d$. The intuitive notion that most of the interaction field energy ought to be "between" the charges is just wrong.

As an aside, I note that negative energy density may raise the specter of wormholes and other esoterica in general relativity[19]. However, the total field energy density is always positive (or zero) since it is proportional to the square of the net field, $(\vec{E}_a + \vec{E}_b)^2$ for two charges. It is the interaction part of the field energy that can be positive or negative.

## 3. More Complex Charge Distributions

The argument presented above about the interaction field energy can be extended to more complex systems of point charges. For example, with three point charges the electric field energy density is proportional to

$$(\vec{E}_a + \vec{E}_b + \vec{E}_c)^2 = E_a^2 + E_b^2 + E_c^2 + 2\vec{E}_a \cdot \vec{E}_b + 2\vec{E}_a \cdot \vec{E}_c + 2\vec{E}_c \cdot \vec{E}_b. \tag{10}$$

In this case, there are three interaction field terms, each of which can be related to a potential energy function. For example, the *bc* term, when integrated over all space, yields

$$W_{bc} = k\frac{q_b q_c}{d_{bc}}, \tag{11}$$

where $d_{bc}$ is the distance between charges *b* and *c*. Thus, the interaction field energy for any static system of point charges is equivalent to the sum of the pair-wise electrostatic potential energy functions. The total electrostatic potential energy can be written as

$$W = \frac{1}{2}k \sum_{i,j\, i\neq j} \frac{q_i q_j}{d_{ij}}, \tag{12}$$

where the factor of ½ accounts for "double counting" of the pairs *i* and *j*.

To apply the potential energy function, one needs to know the distances between the charges and that may be difficult for "polarizable" systems, in which the distances between the charges are affected by all the other charges. Of course, that same effect makes the direct calculation of electric field energy more difficult as well.

## 4. Other Complications with Field Energy

To re-enforce the notion that field energy has many subtleties, I mention a few further complications. First, I note that magnetic field energy is important in plasma physics, as applied, for example to the Sun's corona and to the solar wind (of charged particles) interacting with Earth's magnetic field. The motions of the charged particles and the magnetic field (most often represented as magnetic field lines) are linked in very complex ways[20].

Plesset and Venezian[21] point out that it is easy to confuse the "free energy" associated with electric and magnetic fields (the part that can be translated into work) with the total field energy. The distinction is not necessary for electrostatics and magnetostatics but it does become important when the systems are dynamic.

Potential energy and field energy changes are not always the same, even for static and quasi-static situations. For example, Goedecke, Wood, and Nachman[22] examine the energy associated with the orientation of a magnetic dipole in a magnetic field in terms of potential energy and magnetic field energy and find that those concepts are different.

By considering the physics of a rail gun, Hecking[23] concludes that energy stored in a magnetic field can sometimes be thought of as a form of kinetic energy and in other cases as a form of potential energy.